\documentclass{article}
\usepackage{graphicx}
\usepackage{amsmath,amssymb}

\usepackage{latexsym}
\begin{document}
\title{Model of collimated jets with high energy particles\\
{\Large C. Barbachoux$^{1,2}$\thanks{cecile.barbachoux@obspm.fr}, J. Gariel$^1$\thanks{%
jerome.gariel@upmc.fr}, G. Marcilhacy$^{1}$\thanks{%
gmarcilhacy@hotmail.com} and N. O. Santos$^{1,3}$\thanks{%
nilton.santos@upmc.fr and N.O.Santos@qmul.ac.uk}}\\
{\small
$^1$ LERMA-UPMC, Universit\'e Pierre et Marie Curie,
Observatoire de Paris, CNRS, UMR 8112, 3 rue Galil\'ee, Ivry sur
Seine 94200, France.\\
$^2$ Universit\'e de Nice Sophia Antipolis, Institut Non Lin\'eaire de Nice, UMR CNRS 6618,
1361 route des Lucioles, 06 560 Valbonne, France.\\
$^3$ School of Mathematical Sciences, Queen Mary,
University of London, Mile End Road, London E1 4NS, U. K.}
}

\date{}
\maketitle
\begin{abstract}
\noindent {{\bf Context}The increasing data set of precise observations of
very energetic and collimated jets, with black hole (BH) as putative central
engine, at different astrophysical scales and in various environments,
should soon permit to discriminate and classify current theoritical models
able to describe the jets formation.}\\
{{\bf Aims} Constructing a purely gravitational theoretical model of
perfectly collimated jets of high energy particles in the ideal case where
the central engine is a Kerr BH of mass $M$ and angular momentum by unit of
mass $a$.}\\
{{\bf Methods} Studying in Weyl coordinates ($\rho $, $z$) the unbound
Kerr 2D-geodesics which are asymptotes to straight lines parallel to the $z$
axis of equations
\begin{equation*}
\rho =\mbox{constant}\equiv \rho _{1}= \left[\left(\frac{a}{M}\right)^2+%
\frac{\mathcal{Q}}{E^{2}-1}\right]^{1/2},  \nonumber
\end{equation*}
of which existence was recently demonstrated (Gariel et al.,2010). On these
geodesics, flow test particles of energy $E$, with a Carter constant $%
\mathcal{Q}$ and (necessarily) an angular momentum $L_{z}=0$.}\\
{{\bf Results} We express the motion constants $E$ and $\mathcal{Q}$ as
functions of $r_{1}$ and $r_{2}$, which are real roots of characteristics of the
geodesics equations system. In the special case of a double root $%
r_1=r_2=Y$, and, as an example, fixing $a=M/2$, the $Y$ parametrization of
the constants $E$ and $\rho _1$ displays the following properties: 1) When $%
E\rightarrow \infty $ it implies $\left\vert \mathcal{Q} \right\vert
\rightarrow \infty $, but $\rho _{1}$ remains finite and tends to only
two possible values, $\rho_1/\rho_e =10.241$ and $0.69$ with $\rho_e=a/M$. 2) $E$ \ steeply
decreases from infinity to small values, while $\rho_1$ concomitantly varies little inside two narrow ranges of $\rho_1/\rho_e$: $[10.241;10.65]$ and $[0.69;0.67]$. Thus, the jet has a radial structure. Hence, the energy flux can be
calculated. Furthermore, based on observed data of the jets powers, we can
obtain the mean particles density, the particles flow, the speed and the Lorentz factor
of the jets, for any charged or neutral test particle. Then, we
numerically apply these results to electrons. By studying the
characteristics, we discuss the domains of initial conditions for geodesics
starting inside the ergosphere. All these results come from the Kerr
spacetime structure, and enhances the Penrose process as a plausible
origin for the high energy jets.}{}\\

\end{abstract}

{\bf keywords}:Astrophysical jets; Kerr black hole geodesics; High
energy cosmic rays

\section{Introduction}

Among the numerous observed outflows ejected from various astrophysical
structures at all scales \cite{Levinson,Goveia} the longest, the most
energetical and collimated jets emerge either very fugaciously and intensely
from cosmological sources - like the long duration gamma ray bursts \cite{Aielli} -, or with lifetimes at our time scale, steadily or repeatedly,
nearer, from some microquasars, or, at much larger time and space scales,
from the active galaxies nuclei \cite{Mirabel}. All of them are
relativistic, and often ultra relativistic \cite{Konigl}. Most of these jets
are believed to be powered by a central engine, being a neutron star or BH,
fed by larger structures, like elliptical galaxies, giant companion stars
and X-ray binaries \cite{Russel}. The more and more numerous and precise
\cite{Muller} observations of these various jets will soon help to strongly
constraint the various conjectured theoretical models. In this
perspective, we are here interested to suggest a purely gravitational
theoretical model of the formation of a highly energetic and collimated jet
powered by a rotating BH.

As a first approximation, assuming the system to be axisymmetrical and
in stationary rotation, we can represent a jet as a set of test particles
following Kerr's unbound geodesics focusing at infinity along the $z$ axis.

In this ideal framework, if we succeed to describe a perfectly
collimated jet with high energies, the model will allow us to build
far more realistic descriptions by taking the ambient medium, i.e. the
matter (including a magnetic field), into account.
However, these more realistic descriptions, although
important, will only produce marginal improvements concerning the
origin of the jet formation. The essential focus of the phenomenon will
remain the coherent set of parallel unbound geodesics of a Kerr BH
combined with the source of powering essentially being the gravitational field in its
strongest manifestations, namely the BH.

The generalized cylindrical, or Weyl, coordinates ($\rho $, $z$, $\phi $),
related to Boyer-Lindquist generalized spherical coordinates ($r$, $%
\theta $, $\phi $) by
\begin{equation}
\rho =[(r-1)^{2}-A]^{1/2}\sin \theta ,\;\;z=(r-1)\cos \theta,
\label{eq1}
\end{equation}%
where
\begin{equation}
A=1-\left( \frac{a}{M}\right) ^{2},  \label{1b}
\end{equation}%
are the most suitable for describing observable phenomena generated by
axisymmetric structures. The existence of special unbound geodesics was recently demonstrated in this
framework \cite{Gariel}. These geodesics stem from
the ergosphere and, when $z\rightarrow \infty $, they are asymptotically
parallel to the $z$ axis\textbf{,} with
\begin{equation}
\rho =\rho _{1}\equiv \left( \rho _{e}^{2}+\frac{\mathcal{Q}}{E^{2}-1}\right)
^{1/2}  \label{1a}
\end{equation}%
for the asymptotes, depending on $\rho _{e}\equiv a/M$ and on the two
constants of motion, the Carter constant $\mathcal{Q}$ and the energy $E$,
the third constant of motion, the $z$ component of the angular momentum, $%
L_{z}$ being necessarily null. In the present paper, we
show that only some of these geodesics, belonging to narrow ranges, can be
followed by particles with high energies.

The function $R(r)$ \cite{Chandrasekhar} introduced in the expression of the
Kerr timelike geodesics (test particle mass $\sqrt{\delta _{1}}=1$) plays a
fundamental role in the analysis of the jet collimation in the case of a model
where the engine at the centre of the accretion disk is supposed to be a
stationary rotating BH. This function is such that
\begin{equation}
R^{2}(r)=a_{4}r^{4}+a_{3}r^{3}+a_{2}r^{2}+a_{1}r+a_{0},  \label{1}
\end{equation}%
with (see (2) and (4-8) in \cite{Gariel})
\begin{eqnarray}
a_{0} &=&-a^{2}\mathcal{Q},\;\;a_{1}=2(a^{2}E^{2}+\mathcal{Q}),  \nonumber \\
a_{2} &=&a^{2}(E^{2}-1)-\mathcal{Q},\;\;a_{3}=2,\;\;a_{4}=E^{2}-1,  \label{2}
\end{eqnarray}%
where we put $M=1$ and $L_{z}=0$, considering the special %
2D-geodesics of (\ref{1a})-type. Hence, the BH spin $a$ being fixed, $
-1\leq a\leq 1$, we have two independent parameters left, $\mathcal{Q}$ and $%
E$, or, equivalently from (\ref{1a}), the position $\rho _{1}$
of the asymptote parallel to the $z$-axis and the energy $E$.

The paper is organized as follows. In section 2, we obtain the expressions
of the two motion constants, $E$ and $\mathcal{Q}$, as functions of two real
roots of the characteristics equation $R^{2}(r)=0$. In section 3,
we consider the special case of a double root $Y$, and
we show that there exist only two narrow ranges
of $Y$ for which $E$ can have high energy values. In
section 4, we show that the two possible other roots are functions only of
the two first ones. Then we look at some consequences on the admissible
values of $E$ and the corresponding ranges for the asymptotes $\rho _{1}$.
In section 5,
we calculate the energy flux of the jet, and based on the
observational evaluation of the power of the jet we deduce the corresponding
particles density, the particles flow, the mean velocity and the mean
Lorentz factor of the jet. As an example, we give a numerical estimation of
these quantities for electrons. In section 6,
by studying the characteristics, we show that among the two
previous possibilities found in
section 3, there remains one admissible only. In section 7, as a conclusion,
we discuss qualitatively some potential consequences of relaxing some
restrictive assumptions made here on the possibilities offered by
the Penrose process to obtain high energies with efficient jets formation.

\section{Conserved quantities as functions of two roots}

Let us consider the possible roots of the equation $R^{2}(r)=0$ of the
characteristics $\dot{r}=0$ of the autonomous system of geodesics equations
\cite{Chandrasekhar}, i.e.
\begin{equation}
a_{4}r^{4}+a_{3}r^{3}+a_{2}r^{2}+a_{1}r+a_{0}=0.  \label{5}
\end{equation}%
The polynomial equation (\ref{5}) has four roots, labeled $r_{i}$ with $%
i=1,2,3,4$, which can be \textit{a priori} $\geq 0$ or $\leq 0$ or complex
(contrarily to the $r$ physical variable which is real defined in the range $%
\left[ 1+\sqrt{A},\infty \right[ $). The two equations $R^{2}(r_{1})=0$ and $%
R^{2}(r_{2})=0$ are linear in $\mathcal{Q}$ and in $E^{2}-1$. Solving the
linear system of these two equations yields the two parameters as functions
of the roots $r_{1}$ and $r_{2}$,
\begin{eqnarray}
\mathcal{Q} &=&\frac{2r_{1}r_{2}}{D}\left\{ a^{4}+a^{2}\left[
r_{1}(r_{1}-2)+r_{2}(r_{2}-2)\right] \right.\label{6}\\
&\,&\vspace*{7mm}\left. +r_{1}^{2}r_{2}^{2}\right\},  \nonumber
\\
E^{2}-1 &=&-\frac{2}{D}\left\{ a^{4}+a^{2}(r_{1}^{2}+r_{2}^{2})\right.\label{7}\\
&\,&\vspace*{7mm}\left.+r_{1}r_{2}%
\left[ r_{1}(r_{2}-2)-2r_{2}\right] \right\}, \nonumber
\end{eqnarray}%
with
\begin{eqnarray}
D &=&a^{4}(2+r_{1}+r_{2})+a^{2}\left[
r_{1}^{3}+r_{1}^{2}r_{2}+r_{1}r_{2}(r_{2}-4)+r_{2}^{3}\right]  \nonumber \\
&&+r_{1}r_{2}\left[ (r_{1}^{2}+r_{1}r_{2})(r_{2}-2)-2r_{2}^{2}\right] .
\end{eqnarray}

In the third possible equation, $R^{2}(r_{3})=0$, the parameters $\mathcal{Q}
$ and $E^{2}-1$ can be replaced by (\ref{6}) and (\ref{7}), leading to a
relation between $r_{3}$ and $r_{1}$ and $r_{2}$ allowing, in principle, to
determine the values of $r_{3}$ as function\textbf{s} of $r_{1}$
and $r_{2}$ only, with $a$ being fixed. The fourth possible equation, $%
R^{2}(r_{4})=0$, will not bring any new result because the roots $r_{3}$ and
$r_{4}$ are the same.

In (\ref{6}) and (\ref{7}), it is worth noting the symmetric role of $r_{1}$
and $r_{2}$, and that ${\mathcal{Q}}$ and $E^{2}-1$ have the same
denominator $D$, so that if, and only if, it cancels, we have $E\rightarrow
\infty $ and $\left\vert {\mathcal{Q}}\right\vert \rightarrow \infty $,
whereas $\rho _{1}$, depending only on
their ratio (see (\ref{1a})), tends towards a finite
value. From (\ref{1a}),(\ref{6}) and (\ref{7}) we obtain the asymptotes
\begin{equation}
\left( \frac{\rho _{1}}{\rho _{e}}\right) ^{2}=\frac{%
(a^{2}+r_{1}^{2})(a^{2}-r_{1}r_{2})(a^{2}+r_{2}^{2})}{a^{2}\left\{
a^{4}+a^{2}(r_{1}^{2}+r_{2}^{2})+r_{1}r_{2}\left[ r_{1}(r_{2}-2)-2r_{2}%
\right] \right\} }.  \label{8}
\end{equation}

\section{Roots $r_1=r_2$ real}

For sake of simplification, we assume in this paper that there is a
double real root $r_{1}=r_{2}=Y$. Hence (\ref{1}) can be rewritten as
\begin{equation}
R^{2}(r)=a_{4}(r-Y)^{2}(r^{2}+Br+C),  \label{9}
\end{equation}%
and (\ref{6}) and (\ref{7}) simplify to
\begin{eqnarray}
{\mathcal{Q}} &=&\frac{[a^{4}+2a^{2}(Y-2)Y+Y^{4}]Y^{2}}{%
a^{4}(1+Y)+2a^{2}(Y-1)Y^{2}+(Y-3)Y^{4}},  \label{10} \\
E^{2}-1 &=&-\frac{a^{4}+2a^{2}Y^{2}+(Y-4)Y^{3}}{%
a^{4}(1+Y)+2a^{2}(Y-1)Y^{2}+(Y-3)Y^{4}}.  \label{11}
\end{eqnarray}%
When $E\rightarrow \infty $, of course, $\left\vert Q\right\vert \rightarrow
\infty $ again. But, as already noted, their ratio tends towards a finite
value, so that $\rho _{1}$ remains finite,
\begin{equation}
\left( \frac{\rho _{1}}{\rho _{e}}\right) ^{2}=\frac{%
(a^{2}-Y^{2})(a^{2}+Y^{2})^{2}}{a^{2}[a^{4}+2a^{2}Y^{2}+(Y-4)Y^{3}]}.
\label{12}
\end{equation}

\begin{figure}[ht]
\centering
\includegraphics[width=8cm]{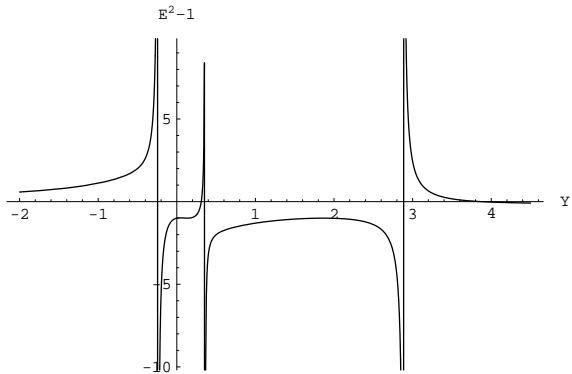}
\caption{Plot of $E^{2}-1=F(Y)$, where $E$ is the energy of the
test-particle, in function of the double root $Y$, evaluated from the
relation (\ref{11}), for a BH of mass $M=1$ and of angular momentum by unit of
mass ${a}/{M}=0.5$. We can see the ranges of $Y$ for which $E^{2}-1$
is positive, as expected for unbound geodesics, and the three values of $%
Y$ for which $E^{2}-1$ tends to the positive infinity.
}
\label{figure1}
\end{figure}

\begin{figure}[ht]
\centering
\includegraphics[width=8cm]{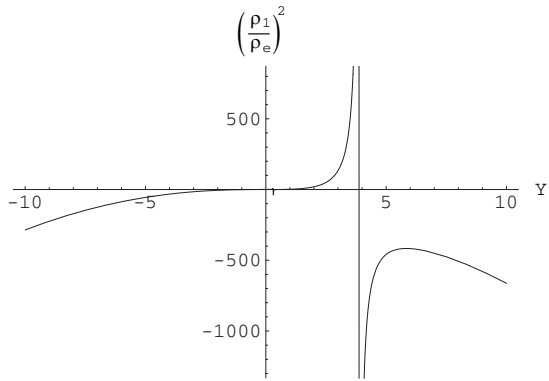}
\caption{Plot of $\left({\rho _{1}}/{\rho _{e}}\right)^{2}=G(Y)$ as function
of the double root $Y$, evaluated from the relation (\ref{12}), for a BH of mass $%
M=1$ and of angular momentum by unit of mass ${a}/{M}=0.5$. As in figure \ref{figure1}, we can see the ranges of $Y$ for which the function $\left({\rho _{1}}/{\rho _{e}}%
\right)^{2}$ is positive.
}
\label{figure2}
\end{figure}

Choosing a "moderate" rotation of the BH, we fix $a=M/2$ (as considered by
us in \cite{Gariel}). Hence, we can plot the functions $E^{2}-1=F(Y)$ and $%
(\rho _{1}/\rho _{e})^{2}=G(Y)$, as shown in figures \ref{figure1} and \ref{figure2}.

Since $F$ and $%
G$ have to be simultaneously $\geq 0$, the only possible solutions
correspond to the two ranges
\begin{eqnarray}
Y&\in &\left[ -0.5,Y_{0a}\right] ,  \label{13} \\
Y&\in &\left[ Y_{0b},3.86971\right] ,  \label{13a}
\end{eqnarray}%
with $Y_{0a}$\ and $Y_{0b}$\ the asymptotes of $F(Y)$, for which $%
E\rightarrow \infty $. We can numerically evaluate these asymptotes (namely the roots
of the equation $D=0$), yielding approximately $Y_{0a}\simeq -0.241806$\ and
$Y_{0b}\simeq 2.8832$ (for more precise values see (\ref{22}) and (\ref{eqY}) respectively).

Hence, there are two only possible values of $\rho _{1}$ for which $%
E\rightarrow \infty $, i.e. one for each range (\ref{13}) and (\ref{13a}). For the limits
of the two ranges (\ref{13}) and (\ref{13a}), $Y=Y_{0a}-\varepsilon $
and $Y=Y_{0b}+\varepsilon $, when $\varepsilon \rightarrow 0$, we obtain the
finite values, respectively,
\begin{equation}
\frac{\rho _{1}}{\rho _{e}}\simeq 0.693199,\;\;\mbox{and}\;\;\frac{\rho _{1}%
}{\rho _{e}}\simeq 10.2411.  \label{14}
\end{equation}

At the other extremity of the range (\ref{13a}),  i.e. for $%
Y=3.86971$ where $\rho _{1}\rightarrow \infty $, we have $E^{2}-1=0$. And for the
other extremity of the range (\ref{13}), i.e. for $Y=-0.5$ where $E^{2}-1=2$, we have $\rho _{1}=0$.

\begin{figure}[ht]
\centering
\includegraphics[width=11cm]{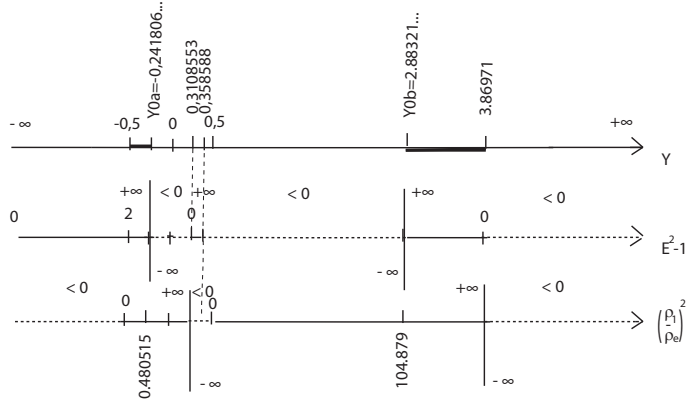}
\label{figure2a}
\caption{Some precise key values of the functions $F(Y)=E^2-1$ and $G(Y)=(\rho_1/\rho_e)^2$ showing the ranges of $Y$ for which these functions are simultaneously positive (bold lines). The negative values of these functions are in dotted lines.}
\end{figure}

The figure \ref{figure2a} summarizes these results.

\section{Roots $r_3$ and $r_4$}

Identifying (\ref{9}) with (\ref{1}) rewritten with the parameters $%
E^2-1=F(Y)$ and $(\rho_1/\rho_e)^2=G(Y)$, without the explicit form of these
functions of $Y$, given by (\ref{11}) and (\ref{12}), yields the four
relations,
\begin{eqnarray}
B-2Y=\frac{2}{F}, \;\; G+\frac{1}{F}=2(BY^2-2YC),  \nonumber \\
1-G=16CY^2, \;\; 2-G=4(C-2YB+Y^2),  \label{16}
\end{eqnarray}
linear in $1/F$, $G$, $B$ and $C$.

After eliminating $1/F$ and $G$ in (\ref{16}), we obtain $B(Y)$ and $C(Y)$,
namely
\begin{eqnarray}
B&=&-\frac{2(4Y^2-1)[Y(4Y-1)+Y-1]}{(4Y^2-1)^2-16Y^2(4Y-1)},  \label{17} \\
C&=&\frac{(4Y^2-1)^2+16Y(Y-1)}{4[(4Y^2-1)^2-16Y^2(4Y-1)]}.  \label{18}
\end{eqnarray}
Hence, (\ref{9}) can be rewritten as
\begin{equation}
R^2=a_4(r-Y)^2(r^2-Sr+P)=a_4(r-Y)^2(r-r_3)(r-r_4),  \label{19}
\end{equation}
where $r_3$ and $r_4$ are the remaining roots, in general distinct, and
\begin{equation}
S\equiv r_3+r_4=-B, \;\; P\equiv r_3r_4=C,  \label{20}
\end{equation}
or
\begin{eqnarray}
r_3&=&-\frac{1}{2}\left[B+\left(B^{2}-4C\right)^{{1}/{2}}\right],\label{21}\\
 r_4&=&-\frac{1}{2}\left[B-\left(B^{2}-4C\right)^{{1}/{2}}\right],  \label{21a}
\end{eqnarray}
where $B(Y)$ and $C(Y)$ are given by (\ref{17}) and (\ref{18}).

\begin{figure}[ht]
\centering
\includegraphics[width=8cm]{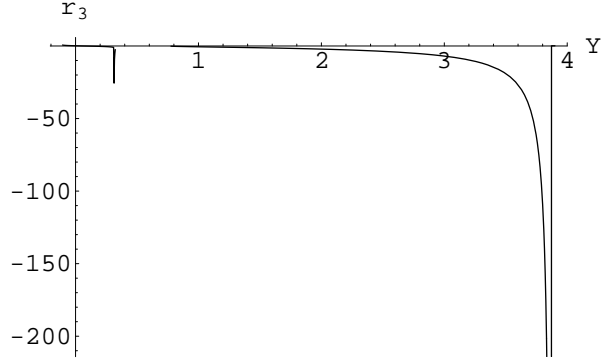}
\caption{Plot of the root $r_{3}$ in function of the double root $Y$,
evaluated from the relation (\ref{21}), for a BH of mass $M=1$ and of angular
momentum by unit of mass ${a}/{M}=0.5$.
}
\label{figure3}
\end{figure}

\begin{figure}[ht]
\centering
\includegraphics[width=8cm]{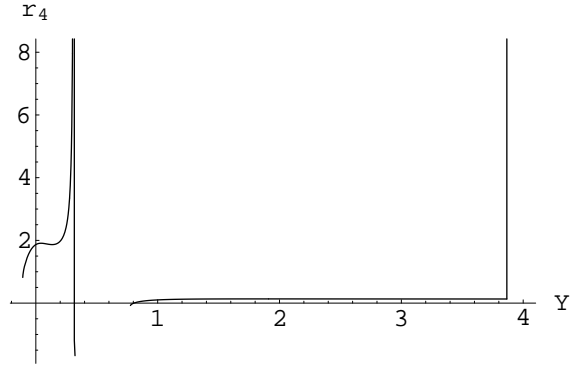}
\caption{Plot of the root $r_{4}$ in function of the double root $Y$,
evaluated from the relation (\ref{21a}), for a BH of mass $M=1$ and of angular
momentum by unit of mass ${a}/{M}=0.5$.
}
\label{figure4}
\end{figure}

The curves $r_{3}(Y)$ and $r_{4}(Y)$, are plotted in figures \ref{figure3} and \ref{figure4}, and
are real for some ranges of $Y$ only. In particular, in the range (\ref{13}) for $Y$, $r_{3}$ and $r_{4}$ are not real. To have the
expression $r^{2}+Br+C$ in Eq. (\ref{9}) real, where $B$ and $C$ are real, $r_{3}$ and $%
r_{4}$ have to be complex conjugated, i.e. $r_{3}=z=B_{1}+iC_{1}$ and $r_{4}=%
\overline{z}$. Hence, the sign of the expression $%
r^{2}+Br+C=(r+B_{1})^{2}+C_{1}^{2}$ is always positive, and $%
P=C=B_{1}^{2}+C_{1}^{2}\geq 0$ and $S=-B=-2B_{1}\leq 0$.

In the range (\ref{13a}) for $Y$, the two roots $r_{3}$ and $r_{4}$ are
real, $P=C$ is negative (which means two roots of opposite signs) and $B=-S$ is
positive. The portions of curves $r_{3}(Y)$ and $r_{4}(Y)$ can be plotted on
the range (\ref{13a}). The most precise value we can numerically obtain for the left limit (where, in principle, $E\rightarrow
\infty $) of the range (\ref{13a}), is
\begin{equation}
Y_{0b}\simeq 2.8832177419263523927462568785232847,
\label{eqY}
\end{equation}%
allowing us to reach the maximal value $E\simeq 1.1\ast 10^{32}$ "only".
Then, the corresponding real values $r_{3}=-5.898835521038341$ and $%
r_{4}=0.13240003718564175$ are obtained.

Also, it is worth observing that $E$ is steeply decreasing, for a weak
variation $\varepsilon $ of $Y$ from $Y_{0b}$ \ ($\varepsilon >0$) or from $%
Y_{0a}$ ($\varepsilon <0$), while $\rho _{1}$ is weakly increasing for this
same small interval of $Y$. For example, when $Y$ goes from $Y_{0b}$ to $%
2.922$, the energy $E$ is steeply decreasing from $10^{30}$ to $3$, while
the position of the asymptote $\rho _{1}/\rho _{e}$ increases by a small
amount from $10.24$ to $10.68$, which means a big concentration of the most
energetic part (the "spine") of the beam immediately near, at the right hand
side, of $\rho _{1}/\rho _{e}=10.24$. At its left hand side, there is no
more beam produced.

Likewise, for the range (\ref{13}), the energy $E$ is very steeply
decreasing from $\infty$
to $6$. The highest value of $E$ ($E_{max}\simeq 2\ast 10^{30}$) is obtained
for the most precise value numerically obtainable for the right
limit $Y_{0a}$, namely%
\begin{equation}
Y_{0a}\simeq -0.241805810271953623405344008301523644957.
\label{22}
\end{equation}
The corresponding asymptote $\rho _{1}/\rho _{e}$, inside the
ergosphere, decreases very slightly from $0.6932$ to $0.6764$. Here the jet is
yet more concentrated just at the left of $0.6932$. While beyond
its right side, there is no possible beam.

As a result, our model predicts a radial structure of the jet, with a
precise profile for its energy (or speed) distribution of the
particles. A radial morphology has already been suggested from observations
\cite{Giroletti,Keppens}.

\section{Flux, particles density and Lorentz factor}

In the region $z$ $\gg \rho_1$ of the jet, where the beam is quasi-parallel
to the $z$ axis, the vector density of (total) energy current is purely
convective, $\vec{j}_E=\rho_E\vec{v}$, where $\rho_E=n\bar{E}$, being $\bar{E%
}$ the mean energy of a particle and $n$ is the number density, $\vec{v}=v%
\vec{e}_z$ its velocity. The power of the jet or, equivalently, the energy
flux across a crown, cross-section of surface $S$, included between two
radii $\rho_{10}$ and $\rho_{11}$, is
\begin{eqnarray}
\mathcal{P}&=&\frac{d\mathcal{E}}{dt}=\int \int_{S}\vec{j}_E\cdot d\vec{S}%
=2\pi \int_{\bar{\rho}_{10}}^{\bar{\rho}_{11}} nv\bar{E}\bar{\rho}_1d\bar{%
\rho}_1\nonumber\\
&=&\pi n\int_{Y_0}^{Y_1}v\bar{E}\frac{d\bar{\rho}_1^2}{dY}\;dY,
\label{23}
\end{eqnarray}
where $d\vec{S}=dS\vec{e}_z$ and we supposed an homogeneous jet, $n=$
constant. The speed of each particle is (see equation (49) of \cite{Gariel}%
),
\begin{equation}
v=c\left(1-\frac{c^4\delta_1}{\bar{E}^2}\right)^{1/2}=\frac{c} {\bar{E}}(%
\bar{E}^2-c^4\delta_1)^{1/2},  \label{24}
\end{equation}
its kinetic energy is defined by
\begin{equation}
\bar{E}_C=(\bar{E}^2-c^4\delta_1)^{1/2},  \label{25}
\end{equation}
and the total energy $\bar{E}$ is linked to the reduced function $E(Y)$,
given in (\ref{11}), by
\begin{equation}
E=\frac{\bar{E}}{c^{2}\sqrt{\delta_1}}.  \label{26}
\end{equation}
Hence we have
\begin{equation}
v\bar{E}=c^3(E^2-1)^{1/2}\sqrt{\delta_1}.  \label{27}
\end{equation}
Likewise, the "true" length $\bar{\rho}_{1}$ in function of the reduced
coordinate $\rho _{1}$, itself function of $Y$ by the equation (\ref{12}),
is
\begin{equation}
\rho _{1}=\frac{\bar{\rho}_{1}}{M}.  \label{28}
\end{equation}%
Whence,
\begin{eqnarray}
\mathcal{P}&=&\frac{d\mathcal{E}}{dt}\nonumber\\
&=&\pi nM^{2}\rho _{e}^{2}c^{3}\sqrt{\delta
_{1}}\int_{Y_{0}}^{Y_{1}}(E^{2}-1)^{1/2}\frac{d(\rho _{1}/\rho _{e})^{2}}{dY}%
\;dY,  \label{29}
\end{eqnarray}%
where the adimensional functions $E(Y)$ and $(\rho _{1}/\rho _{e})^{2}(Y)$
are given by the equations (\ref{11}) and (\ref{12}). The relation (\ref{29}%
) can be used to calculate, for example, the power of the beam of the
particles ejected along the geodesics with asymptotes included between $\rho
_{10}/\rho _{e}=0.69319914385$ and $\rho _{11}/\rho _{e}=0.676252$,
corresponding, as seen in section 4, to energies varying from the infinity
(in fact, $E\sim 2\ast 10^{32}$) to $5.8$, and with $Y$ range (\ref{13}) from $%
Y_{0a}\simeq -0.241806$ to $Y_{1}=-0.250$. However this calculation would suppose
knowing the particles density $n$ inside the jet. Conversely, if the power
(which is an observational data more easily attainable by other independent
methods \cite{Willott,Fender,Punslya,Punslyb}), is known, then from our
model we can deduce the density $n$.

Further assumptions have to be made about the ejected particles and the BH
parameters. Let us consider a supermassive BH with mass $M=10^{9}M_{\odot }$
(with the solar mass $M_{\odot }=3$km) and angular momentum by unit mass $%
a=M/2$ , and electrons with restmass $\sqrt{\delta _{1}}=0.511$MeV. Let us
suppose that the jet power is $\mathcal{P}=10^{47}$ erg/s \cite%
{Willott,Punslya}. We find with this data (the calculation can be made numerically from a series expansion of the integrant near $Y_{0a}$, which is
justified because $\left\vert Y-Y_{0a}\right\vert \leq \left\vert
Y_{1}-Y_{0a}\right\vert \simeq 8.2\ast 10^{-3}<1$),
\begin{equation}
I=\left\vert \int_{Y_{0a}}^{Y_{1}}(E^{2}-1)^{1/2}\frac{d(\rho _{1}/\rho
_{e})^{2}}{dY}\;dY\right\vert =0.264939.  \label{30}
\end{equation}%
Whence the density is deduced
\begin{equation}
n=\frac{10^{47}}{8.29\ast 10^{28}}=2.18\ast 10^{20}\;\mbox{electrons/m}^{3}.
\label{31}
\end{equation}
The flow of particles across the same crown of surface $S$ is
\begin{eqnarray}
\frac{dN}{dt} &=&n\pi \int_{S}vd\rho _{1}^{2}\nonumber\\
&=& n\pi M^{2}\rho
_{e}^{2}c\int_{Y_{0a}}^{Y_{1}}\left( 1-\frac{1}{E^{2}}\right) ^{1/2}\frac{%
d(\rho _{1}/\rho _{e})^{2}}{dY}\;dY  \nonumber \\
&=&n\ast 3.14\ast \frac{27}{4}\ast 10^{32}\ast J,  \label{32}
\end{eqnarray}%
and we obtain,
\begin{eqnarray}
J&=&\left\vert \int_{Y_{0a}}^{Y_{1}}\left( 1-\frac{1}{E^{2}}\right) ^{1/2}%
\frac{d(\rho _{1}/\rho _{e})^{2}}{dY}\;dY\right\vert \nonumber\\
&=&2.30373\ast 10^{-2},
\label{33}
\end{eqnarray}%
and finally,
\begin{equation}
\frac{dN}{dt}=1.06352\ast 10^{52}\;\mbox{electrons/s}.  \label{34}
\end{equation}%
From these results the mean energy by particle can be deduced,
\begin{eqnarray}
&<&\bar{E}>=\frac{d\mathcal{E}}{dt}\left( \frac{dN}{dt}\right) ^{-1}
\nonumber \\
&=&0.511\ast \frac{0.264939}{2.30373\ast 10^{-2}}\simeq 5.87673\;%
\mbox{MeV/electron},  \label{35}
\end{eqnarray}%
which yields the (mean) Lorentz factor
\begin{eqnarray}
\Gamma _{m}\equiv <E>&=&\frac{<\bar{E}>}{c^{2}\sqrt{\delta _{1}}}=\frac{I}{J}=%
\frac{0.264939}{2.30373\ast 10^{-2}}\nonumber\\
&=&11.5005,  \label{36}
\end{eqnarray}%
and the mean velocity
\begin{equation}
v_{m}=c\left( 1-\frac{1}{<E>^{2}}\right) ^{1/2}=0.996212\ast c,  \label{37}
\end{equation}%
corresponding to an average ultra relativistic jet.

Now, we also can consider a narrow, more energetic, part of the jet only. As
examples, let us consider the part included between $\rho _{10}$ and $\rho
_{12}=0.69319\ast \rho _{e}$ (or $\rho _{13}=0.693199\ast \rho _{e}$),
corresponding to $Y_{0a}$ and $Y_{2}=-0.24181$ (or $Y_{3}=-0.241805810272$),
respectively, i.e. to thickness $\delta \rho _{2}=\rho _{12}-\rho
_{10}=-8.68546\ast 10^{-6}\ast \rho _{e}$, for $Y_{2}-Y_{0a}=-4.18973\ast
10^{-6}$ (or $\delta \rho _{3}=\rho _{13}-\rho _{10}=-9.61453\ast
10^{-14}\ast \rho _{e}$, for $Y_{3}-Y_{0a}=-4.63518\ast 10^{-14}$), and
hence to ratio of flux surfaces $S_{2}/S=(\rho _{12}^{2}-\rho
_{10}^{2})/(\rho _{11}^{2}-\rho _{10}^{2})=5.18849\ast 10^{-4}$ (or $%
S_{3}/S=(\rho _{13}^{2}-\rho _{10}^{2})/(\rho _{11}^{2}-\rho
_{10}^{2})=5.74535\ast 10^{-12}$).

Furthermore, let us (reasonably) assume
that the power by unit surface crossed by the jet (i.e. the energetic flux)
is constant:
\begin{equation}
\frac{\mathcal{P}}{S}=\frac{\mathcal{P}_{i}}{S_{i}},\;\;i=2,3.  \label{38}
\end{equation}

Hence, the part of the jet with the thickness $\delta \rho _{2}$ (or $\delta
\rho _{3}$), will have a kinetic power $\mathcal{P}_{2}=(S_{2}/S)\mathcal{P}%
=5.18849\ast 10^{43}$ erg/s (or $\mathcal{P}_{3}=(S_{3}/S)\mathcal{P}%
=5.74534\ast 10^{35}$ erg/s), a particle density $%
n_{2}=(S_{2}/S)(I/I_{2})n=4.97047\ast 10^{18}$ electrons/m$^{3}$ (or $%
n_{3}=(S_{3}/S)(I/I_{3})n=5.23365\ast 10^{14}$ electrons/m$^{3}$), a flow of
particles $dN_{2}/dt=1.26855\ast 10^{47}$ electrons/s (or $%
dN_{3}/dt=1.48031\ast 10^{35}$ electrons/s), a mean energy $<\bar{E}_{2}>=%
\sqrt{\delta _{1}}(I_{2}/J_{2})c^{2}=255.631$ MeV (or $<\bar{E}_{3}>=\sqrt{%
\delta _{1}}(I_{3}/J_{3})c^{2}=2.42573$ TeV), a mean velocity $%
v_{m2}=0.99999800204c$ (or $v_{m3}=0.9999999999999778c$) and a Lorentz
factor $\Gamma _{2}=I_{2}/J_{2}=500.257$ (or $\Gamma
_{3}=I_{3}/J_{3}=4.74703\ast 10^{6}$). Let us precise that $%
I_{2}=-6.02378\ast 10^{-3}$ and $J_{2}=-1.20414\ast 10^{-5}$ (or $%
I_{3}=-6.33485\ast 10^{-7}$ and $J_{3}=-1.33449\ast 10^{-13}$) are the
integrals (\ref{30}) and (\ref{33}) respectively, in which the upper limit becomes $%
Y_{2}$ (or $Y_{3}$) instead of $Y_{1}$. So, we can see that, in its core,
the jet is as more strongly UR as nearer of $\rho _{10}$. Let us precise
that $Y_{3}$, and all the corresponding values with the index "$3$",
corresponds to the value of $Y$ the nearest of $Y_{0a}$ for which we are yet
able to numerically evaluate the integrals $I_{3}$ and $%
J_{3}$ (and so, the different corresponding quantities ). Let us note that
the conditions inside the jet for the most narrow channel we are able to
evaluate are comparable to conditions inside a terrestrial particles
accelerator (e.g. LHC), namely $<\bar{E}_{3}>\sim 2$ TeV inside a channel of
thickness $\sim \delta \bar{\rho}_{3}\simeq 14.4218$ cm (but here for
electrons, while in the LHC there are protons. We can evaluate jets of
protons as well, which would give us yet higher energies: factor $\sim 2\ast
10^{3}$). The results are summarized in the table 1.
\begin{table}[ht]
\begin{minipage}[c]{0.5\linewidth}\centering
\resizebox{12cm}{!} {
\begin{tabular}{||c|c|c|c||}
 \hline
&$ Y_1$&$Y_2$&$Y_3$\\
\hline
$Y$&$-0.250$&$-0.24181$&$-0.241805810272$\\
\hline
$Y-Y_{0a}$&$-8.19419\ast 10^{-3}$&$-4.18973\ast 10^{-6}$&$-4.63518\ast 10^{-14}$\\
\hline
$E$&$5.81378$&$250.13$&$2.37778\ast 10^6$\\
\hline
$ {\delta\rho_i}/{\rho_e}$&$-0.0169469$&$-8.68546\ast 10^{-6}$&$-9.61453\ast 10^{-14}$\\
\hline
$\delta\rho_i\;\;(m)$&$2.54204\ast 10^{10}$&$1.30282\ast 10^7$&$14.4218\ast 10^{-2}$\\
\hline
$P\;(erg/s)$&$10^{47}$&$5.18849\ast 10^{43}$&$5.74534\ast 10^{35}$\\
\hline
$I$&$-0.264939$&$-6.02378\ast 10^{-3}$&$-6.33485\ast 10^{-7}$\\
\hline
$n\; \;(e^-/m^3)$&$-2.18711\ast 10^{20}$&$-4.97047\ast 10^{18}$&$5.23365\ast 10^{14}$\\
\hline
$J$&$-2.30373\ast 10^{-2}$&$-1.20414\ast 10^{-5}$&$-1.33449\ast 10^{-13}$\\
\hline
${dN}/{dt} \;\;(e^-/s)$&$1.06352\ast 10^{52} $&$1.26855\ast 10^{47}$&$1.48031\ast 10^{35}$\\
\hline
$E_m \;\;(Mev)$&$5.87673$&$255.631$&$2.42573\ast 10^6$\\
\hline
$1-v_m/c $&$3.788\ast 10^{-3}$&$1.99796\ast 10^{-6}$&$2.22\ast 10^{-14}$\\
\hline
$\Gamma$ &$11.5005$&$500.257$&$4.74703\ast 10^6$\\
\hline
\end{tabular}
}
\end{minipage}
\caption{Values of the departures of the position from the asymptotes at high energies, the thickness of the jet, its density, its mean energy, its mean velocity and its (mean) Lorentz factor, as function of the parameter $Y$, for $Y$ in the vicinity of $Y_{0a}$ (\ref{22}), where $E\rightarrow \infty$ and ${\rho_1}/{\rho_e}\rightarrow 10.24106$ ($E_{max}=1.1\ast 10^{32}$). The integrals $I$ and $J$ are defined by (\ref{30}) and (\ref{33}) with the corresponding upper limit $Y_i$ $(i=1,2,3)$, the lower limit being $Y_{0a}$.}
\end{table}

As second example, we give the same evaluations for the second part of the
very energetic jet, namely near the value $Y_{0b}\simeq 2.88$ (\ref{eqY}). The results are summarized in the table 2.
We shall not comment more precisely these last results, because they cannot
correspond to effective jets, as we shall show in the next section.

\begin{table}[ht]

\resizebox{12cm}{!} {
\begin{tabular}{||c|c|c|c|c||}
\hline
&$Y_1$&$Y_2$&$Y_3$&$Y_4$\\
\hline
$Y$&$2.922$&$2.833$&$2.88321774193$&$2.883217741926354$\\
\hline
$Y-Y_{0b}$&$3.87823\ast 10^{-2}$&$8.2258\ast 10^{-5}$&$3.6473\ast 10^{-12}$&$1.33227\ast 10^{-15}$\\
\hline
$E$&$3.027$&$63.7027$&$302\,514$&$1.50555\ast 10^7$\\
\hline
$ {\delta\rho_i}/{\rho_e}$&$0.411539$&$8.49897\ast 10^{-4}$&$3.76801\ast 10^{-11}$&$1.77636\ast 10^{-14}$\\
\hline
$\delta\rho_i\;\;(m)$&$6.17309\ast 10^{11}$&$1.27485\ast 10^{9}$&$56.5201$&$2.66454\ast 10^{-2}$\\
\hline
$P\;(erg/s)$&$10^{47}$&$2.02457\ast 10^{44}$&$8.97533\ast 10^{36}$&$3.96648\ast 10^{33}$\\
\hline
$I$&$49.1803$&$2.21766$&$4.668989\ast 10^{-4}$&$7.2287\ast 10^{-6}$\\
\hline
$n\; \;(e^-/m^3)$&$1.17177\ast 10^{18}$&$5.26106\ast 10^{16}$&$1.10759\ast 10^{13}$&$3.16214\ast 10^{11}$\\
\hline
$J$&$8.33589$&$1.74073.\ast 10^{-2}$&$7.62157\ast 10^{-10}$&$3.63748\ast 10^{-12}$\\
\hline
$ {dN}/{dt} \;\;(e^-/s)$&$2.07028\ast 10^{52}$&$1.94106\ast 10^{48}$&$1.78919\ast 10^{37}$&$2.43823\ast 10^{33}$\\
\hline
$E_m \;\;(Mev)$&$3.01481$&$65.1005$&$313\,100$&$1.01536\ast 10^6$\\
\hline
$1-v_m/c$&$1.4469\ast 10^{-2}$&$3.1\ast 10^{-5}$&$1.3318\ast 10^{-12}$&$1.266\ast 10^{-13}$\\
\hline
$\Gamma$ &$5.89983$&$127.398$&$612\,719$&$1.98701\ast 10^{6}$\\
\hline
\end{tabular}
}

\caption{Same values as table 1 but in the vicinity of $Y_{0b}$ (\ref{eqY}) where $E \rightarrow \infty$ ($E_{max}=2\ast 10^{30}$) and $ {\rho_{1}}/{\rho_e}\simeq 0.69$.}
\end{table}

\section{Geodesics for high energy jets}

Now let us look at the possible geodesics framing a jet for values of the
energy and of the corresponding asymptotes $\rho_1$ given in section 5. The
existence of admissible initial conditions for such geodesics are provided
by the existence and positions of the 2D-characteristics of the system of
geodesics equations (see (2) and (3) in \cite{Gariel}). Each characteristics
curve delimits two separated parts (regions) in the plane $(r,\theta)$ as
predicted by the theory \cite{Pontriaguine} often applied to the so-called ``qualitative analysis". Each part contains a set of geodesics, which can never
cross the characteristics towards the other part. In Boyer-Lindquist
coordinates, the characteristics are defined by the equations
\begin{equation}
{\dot r}=0, \;\; {\dot \theta}=0,  \label{39}
\end{equation}
which are equivalent, from equations (19) and (20) in \cite{Gariel}, to the
system of algebraic equations,
\begin{equation}
P=0, \;\; S=0,  \label{40}
\end{equation}
the solutions of which, when they exist, are some values  of $r_i$ of $r$ (previously introduced in Sections (2) to (4)) and some values $\theta_i$ of
$\theta$, respectively (which define circles and straight lines from the origin respectively).

In Weyl coordinates, $\rho $ and $z$, these characteristics equations (\ref{40}) are equivalent to the equations (see (17) and (18) in
\cite{Gariel})
\begin{equation}
{\dot \rho }=\frac{S(\alpha^2-A)z}{\alpha \rho \Delta}, \;\; {\dot z}=-\frac{%
S\alpha }{\Delta},  \label{41}
\end{equation}
and
\begin{equation}
{\dot \rho }=\frac{P\alpha ^{3}\rho }{(\alpha ^{2}-A)\Delta}, \;\; {\dot z}=%
\frac{P\alpha z}{\Delta},  \label{42}
\end{equation}
respectively, where
\begin{equation}
\Delta=(\alpha+1)^{2}\alpha ^{2}+\left(\frac{a}{M}\right)^{2}z^{2}.
\label{42a}
\end{equation}
Each set of equations (\ref{41}) and (\ref{42}) lead to,
\begin{equation}
\frac{dz}{d\rho }=-\frac{\alpha^2\rho}{(\alpha^2-A)z},  \label{43}
\end{equation}
and
\begin{equation}
\frac{dz}{d\rho }=\frac{(\alpha^2-A)z}{\alpha^2\rho },  \label{44}
\end{equation}
respectively, defining the two families of characteristics for the geodesics of
type (21) in \cite{Gariel} in which we are interested, namely, ellipses
(corresponding to ${\dot r}=0$) and hyperboles (corresponding to ${\dot
\theta }=0$). Let us note that the product of the two derivatives (\ref{43})
and (\ref{44}) of these characteristics is $-1$ (which confirms that they are
orthogonal).

The first ones, ellipses, exist when there are solutions $r=r_i=$ constant
of (\ref{43}) for $\forall \theta $, with $r_i\geq 1+\sqrt{A}$ or
equivalently $\alpha =\alpha_i=$ constant ( because $\alpha =r-1$) with $%
\alpha_i\geq \sqrt{A}$. Then (\ref{43}) can be integrated yielding
\begin{equation}
\left(\frac{z}{\alpha_i}\right)^2+\frac{\rho^2}{\alpha_i^2-A}=K_1,
\label{45}
\end{equation}
where $K_1$ is an integration constant. Comparing (\ref{45}) with the equation (12) of
\cite{Gariel} imposes $K_1=1.$

The second ones, hyperboles, exist when there exist solutions $\mu =\mu_i=$
constant of (\ref{44}) for any $r$, with $\mu_i^2\leq 1$. These are
solutions of the equation $S=0$, when $L_z=0$, with
\begin{equation}
S^{2}=\left(\frac{a}{M}\right)^{2}(E^{2}-1)\alpha^4 \left[1-\left(\frac{z}{%
\alpha }\right)^2\right] \left[\frac{\mathcal{Q}}{a^{2}(E^{2}-1)}+\left(%
\frac{z}{\alpha}\right)^2\right].  \label{46}
\end{equation}
There are two possible cases, namely $\mu^2_i=1$, then $S=0$ for  any $\mathcal{Q}$, or $\mu_{i}^{2}=-\mathcal{Q}/[a^2(E^2-1)]
=1-(\rho_1/\rho_e)^{2}\leq 1$ , being positive defined only if $\mathcal{%
Q\leq }0$, or equivalently, if $\rho_1\leq \rho_e$. Then, for $\mu^2_i=1$,
we have $z=r-1=\alpha $ and $\rho =0$ for any $r$ and (\ref{44}) reduces to $%
\rightarrow \infty $\ , and the characteristics being along the semi-axis $%
z\geq \sqrt{A}$. While for $\mu_i^2=-\mathcal{Q}/[a^{2}(E^2-1)]$ we have
for (\ref{44})
\begin{equation}
\frac{dz}{d\rho }=\frac{z^2-A\mu_i^2}{z\rho},  \label{47}
\end{equation}
which can be integrated leading to
\begin{equation}
\rho=K_2\left[\left(\frac{z}{\mu_i}\right)^2-A\right]^{1/2},  \label{48}
\end{equation}
where $K_2$ is an integration constant. Comparing (\ref{48}) with (12) in
\cite{Gariel} we have $K_2+\mu_i^2=1$.

The expression (\ref{48}) represents a family of hyperboles parametrized by
\begin{equation}
\frac{\rho_1}{\rho_e}=\left(1-\mu_i^2\right)^{1/2},
\end{equation}
yielding
\begin{equation}
\frac{1}{A}\left[1-\left(\frac{\rho_1}{\rho_e}\right)^2\right]^{-1}z^2-
\frac{1}{A}\left(\frac{\rho_1}{\rho_e}\right)^{-2}\rho^2=1.  \label{49}
\end{equation}

If the initial condition (IC) of a geodesics lies inside an ellipse of the
type (\ref{45}), this geodesics cannot be an unbounded geodesics, and hence
cannot go to infinity. So, the admissible IC have to satisfy the triple
condition: i) being inside the ergosphere, in order to be possibly issued
from a Penrose process; ii) being outside the larger elliptic
characteristics, this means corresponding to the larger value of the roots $%
r_i$; and iii) being above the higher hyperbolic characteristics, which
corresponds to the higher values of the roots $\left\vert \mu_i\right\vert
\leq 1$. That restricts the admissible domain of IC.

An ellipse (\ref{45}), when it exists (i.e. when $r_i\in \left[1+\sqrt{A}%
,\infty \right[ $), can intersect the ergosphere only if its semi-minor axis
$b_{i}=(\alpha_i^2-A)^{1/2}$ is smaller than $\rho_e=a/M$, i.e. if $r_i<2$.

\begin{figure}[ht]
\centering
\includegraphics[width=8cm]{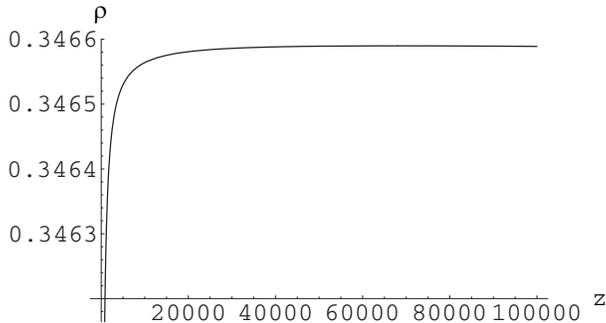}
\caption{Plot of the geodesics $\rho (z)$, for the parameters $%
a=0.5M$ and $M=1$ of the BH,  the motion constants $L_{z}=0$, $E=10^{6}$, $%
\mathcal{Q=}\left[\left(\rho _{1}/\rho _{e}\right)-1\right](E^{2}-1)\left(a/M\right)^2$
(with ${\rho _{1}}/{\rho _{e}}=0.693199$) of the test-particle, and for
the initial conditions $\rho _{i}=2.8\ast 10^{-6},$ $z_{i}=0.852086186870110$%
, inside the ergosphere. We can observe that this geodesics is
asymptotically parallel to the $z$-axis at the asymptote of equation $\rho
=\rho _{1}\equiv 0.3466$, at least until the altitude $z\simeq 8\ast 10^{4}$. For a greater altitude, a greater precision on the initial conditions
would be necessary, theoretically without any limitation.
}
\label{figure5}
\end{figure}

As example, let us take the special case of a double root $Y$ studied in the
precedent sections.

a) The first admissible range that we found is $Y\in \lbrack Y_{0b},3.86]$
(see figure \ref{figure2a}). These roots, belonging to the domain of physical definition,
$r\in \left[ 1+\sqrt{A}=1.86,\infty \right[ $, all correspond to the
existence of elliptic characteristics. The smallest ellipse has as
semi-minor axis $%
b_{i}=[(Y-1)^{2}-A]^{1/2}=[(1.88)^{2}-0.75]^{1/2}=(2.78844)^{1/2}$ along $%
\rho $, and as semi-major axis $a_{i}=\sqrt{\alpha _{i}^{2}}=Y-1=1.88$ along
$z$, obtained for the smallest value $Y_{0b}\simeq 2.88$, corresponding to $%
\rho _{1}/\rho _{e}\simeq 10.24$. This ellipse contains the ergosphere, the
limits of which being $z_{\max }=\sqrt{A}=0.866025$ and $\rho _{\max }=1/2$.
Hence, it is always impossible to have IC simultaneously inside the
ergosphere and outside any ellipse. There is no possibility of unbound
geodesics starting from the ergosphere in this first case.

b) For the second admissible range we found that $Y\in \lbrack -0.5,Y_{0a}]$
(see figure \ref{figure2a}). These roots do not belong to the domain of definition of the
physical variable $r$, which means that there are never any
corresponding elliptic characteristics. The only remaining possible
limitation depends on the position of the hyperbolic characteristics (\ref{49}). The
hyperbola intersects the $z$ -axis at the point with coordinates $\rho =0$
and $z_{0}=\{A[1-(\rho _{1}\rho _{e})^{2}]\}^{1/2}$ and tends asymptotically
towards the straight line of equation $\rho \simeq z\tan \theta _{1}$, with $%
\sin \theta _{1}=\rho _{1}/\rho _{e}$. The domain of possible IC is located
between the $z$-axis, the limit of the ergosphere and above the hyperbola.
For example, for $Y=-0.241806$, $\rho _{1}/\rho _{e}=0.693199$, $\theta
_{1}=21^o$, and $z_{0}=0.624185(<\sqrt{A}=0.8660254037844386)$. We plot in figure \ref{figure5}
the geodesics which tends asymptotically towards the corresponding $\rho
=\rho _{1}=0.3466$, for which the test particle has a very high
(theoretically infinite) energy (for the calculations, we choose the value $%
E=10^{6}$ and $\rho _{1}/\rho _{e}=0.693199$). This plot corresponds to the
IC $\rho _{i}=2.8\ast 10^{-6}$ and $z_{i}=0.852086186870110$ which are
(just) inside the ergosphere at its top near the $z$-axis, i.e. near the
event horizon. For the other limit, $Y=-0.5$, of the $Y$ range, $\rho _{1}=0$
, $E=\sqrt{3}$, $\theta _{1}=0$ and $z_{0}=\sqrt{A}=0.866$.

\section{Discussion}

Taking as parameters the roots of a characteristics equation for unbound
2D-geodesics with $L_{z}=0$, we showed that the two remaining motion
constants, $E$ and $\mathcal{Q}$, of a test particle following geodesics
which asymptotically tends towards a parallel line to the $z$-axis, can be
deduced as a function of these parameters. In the special case of a double
root, and choosing the BH angular momentum by unit of mass $a=M/2$,
restricted domains of $\rho _{1}$ asymptotes corresponding to high energies
are found. That means that the Kerr metric can generate powerful collimated
jets in some precise regions only (and, as a consequence, from some precise
regions of the ergosphere). Indeed, we obtained, in this special
case, two only possible ranges of $\rho _{1}$, namely $\rho _{1}\in \left[
0.3382,0.3466\right] $ and $\rho _{1}\in \left[ 5.12,5.34\right] $ for $E\in %
\left[ \sqrt{6},\infty \right[ $ and $E\in \left] \infty ,\sqrt{3}\right] $
respectively (see the figure \ref{figure2a}). Then,\textbf{\ }for electrons, we calculated
the particle density, the particle flow, the thickness of the jet and the
Lorentz factor, exhibiting UR jets.The numerical evaluations were summarized
in the tables 1 and 2.

These results can be pertinent both for jets and for ultra high energy
cosmic rays (UHECR). Indeed, such energies can be obtained with quasi unlimitedly values
inside the ergosphere thanks to the Penrose process. That
comes from the fact that this process implements three levels (orders of
magnitude) of rest masses (or of energies): the BH mass (which could be
designated as a "supermacroscopic", or astrophysical mass), the incident mass of the body
going into the ergosphere along the equatorial plane from the accretion disk
(which could be called a "macroscopic" mass); and the particle
going out of the ergosphere asymptotically parallel to the $z$ axis (which
could be called a "microscopic" mass).

The "microscopic" outgoing particle, via the original Penrose process \cite{Penrose}, simply by its separation from the incident macroscopic body inside the
ergosphere, can acquire a kinetic energy of order of magnitude of
this ingoing mass, i.e. a "macroscopic" kinetic energy.

The simplest example of such a process could be the ionization of an ingoing
atom of hydrogen inside the ergosphere, with its nucleus irreversibly
falling into the event horizon while the electron is ejected along a
geodesic asymptotically to $\rho_1$ with a kinetic energy increased of about
$1836$ times its rest mass. The rest mass of the nucleus (the proton) could
be considered as the lowest limit of a macroscopic body, being in this case
the electron the microscopic particle.

If the principle of this phenomenon, for the formation of UHECR in
collimated jets, is realistic, we have to expect the total mass of the
ingoing body to be "macroscopic" as compared to (i.e. infinitely larger
than) the total mass of the ejected particles, and to be of the same order
of magnitude than the mass falling in the BH after each Penrose type decay
inside the ergosphere.

Then, in terms of "rest mass" (i.e. of the amount of matter), the efficiency
of the Penrose process is very (infinitely) weak, meaning that a very weak
part of the ingoing mass is ejected to the outside.
But in terms of energy
(a conserved quantity), the Penrose process is very (infinitely) efficient,
meaning that the energy of the ejected particles is much (infinitely) larger
than the energy of these same incident microscopic particles (when they were
linked yet to the incident macroscopic body).

We can make an analogy with a "superelastic" shock, as sometimes occurs in
nuclear physics, in which the total kinetic energy of the\ final outgoing
particles is greater than the kinetic energy of the \textbf{\ }initial
ingoing particles, the difference being acquired to the detriment of a part
of the ingoing (rest) mass, called "mass defect" (as opposed to an
"elastic shock", for which the kinetic energy is conserved, or to an
"inelastic shock" for which the final kinetic energy is lower than the
initial one). In our case, the lost mass falling into the BH is equivalent
to the "mass defect", and is found again in the form of kinetic energy of
the ejected mass. Besides, the efficiency of a collimated jet formation (in
terms of mass as well as of energy) is apparently even more weakened by the fact
that, on the set of ejected particles (in all the directions from the
ergosphere, by this process), very few are ejected in a direction which is
asymptotically parallel to the $z$-axis \cite{Metzger}. However, the only possible
collimation being in this direction, the coherent set of these particles
gives them a privileged observable character, which does not exist for the
other particles isolatedly scattered in all the directions.

Recent results of the Pierre Auger Observatory \cite{Roulet} seem to show a
correlation between the UHECR (above 57EeV) and the nearby ($<71$ Mpc))
AGNs. It would be interesting to examine more precisely the correlations
with the directions (radioloud, quasars or blazars) of the jets and with the
proximity of their sources.

All the results, up to the section 4 (in particular the numerical
values, which concern adimensional quantities), are strict consequences of
the structure of the Kerr metric where we solely fixed the BH spin, $a=M/2$,
and we made the assumption of a double real root $Y$.

A first expected position of the asymptote corresponding to an
"infinite" energy in our model is $\rho_1\simeq 10\rho_e=5M=2.5\ast r_S$.
The second asymptote predicted by our model for a very high energy is $\rho_1=(0.693199/2)M\simeq 0.3466M$, i.e. at about $1/6$
of the Schwarzschild radius $r_S=2M$. This last case is the only compatible with
the limitations imposed by the characteristics of the system of geodesics
equations, as seen in section 6. That means that the presently described jet is a
very thin (narrow) jet.

Besides, there will be a strong concentration of the most energetic part of
the beam in the close vicinity of these two asymptotes, with, on one side, an
abrupt (steep) decreasing and, on the other side, a smoother decreasing, which
indicate a radial energetic structure of the jet, as observed \cite%
{Giroletti}.

The rarity of the currently detected UHECR \cite{Auger} could be explained
in our model by the rarity of UHE particles in the very narrow beams near
the two asymptotes shaping the jet, i.e. directly observable in the very
precise axial direction only.
Even though the energies have to be limited ($%
<50EeV$) by the Greisen-Zatsepin-Kuzmin effect, firstly primary particles
with energies $<50EeV$ could come from nearby ($<70\ MPc$) AGN to the
earth (principally neutrinos), and secondly, for more energetic and more
distant primary particles, a  particular mechanism, for instance with secondary
particles, can be considered \cite{Essey}. Anyway, the recent observations
of multi-TeV photons from distant blazars which do not display the expected
spectrum (with suppressions due to the interaction with the cosmic microwave
background) require an explanation \cite{Aharonian}.

Let us briefly recall that we found these results under some restrictive
conditions, taken into account in section 3, especially the assumption of the existence of a double root $Y$,
and the choice of the "middle" value $a=M/2$. We can thus
hope that by "relaxing" these assumptions, other more general results could
emerge.

For instance, relaxing the assumption of a double root $Y$, while keeping
the same value $M/2$ of the parameter $a$, could open the possibility of
other solutions with $\rho_1>\rho_e$, i.e. thicker jets, for admissible high
energy jets. Work is in progress in that way and our first results are
encouraging. Another example could be to study, in this model, the role of
the BH spin $a$ by trying (say) higher values of this parameter. Let us
note, by the way, that there are few observational results concerning the
possible values of $a$ \cite{Zhang,Istomin,Aschenbach,Aschenbach1}, and its
possible correlation with the length and power of the jet, while there are
more numerous observations concerning the values of $M$.

Our results can also easily be extended to particles other
than electrons, for example to protons or neutrinos. This does not change
the "geometry" that we obtained, i.e. the positions $\rho _{1}$ of the jets,
but their energy only.

For a proton ($\sqrt{\delta _{1}}\simeq 1GeV$) the
maximal energy we can here numerically calculate (but which is
theoretically as large as we want) is about $E\simeq 5.6\ast
10^{25}eV=5.6\ast 10^{7}EeV$, which largely includes the highest energies
of the current observed UHECR \cite{Auger,Dermer,Hoover}.

Finally, let us also recall that our model does not require magnetic fields,
which allows us to discard some problems related to their strength \cite{deSouza}, necessary to obtain such huge energies, and permits us to
consider neutral particles as well. For example, a neutrino, which mass is assumed to be
$\sqrt{\delta_1}=0.33 eV$ \cite{Steidl}, for the precedent evaluation,
would reach the energy $E\simeq 2\ast 10^{-2} EeV$, which seems to be an
acceptable value \cite{Gorham,Hoover,Berezinsky}.

In our model, the only role played by a magnetic field is the role of the
induced magnetic field which tends to stabilize the jet (e.g. \cite%
{Appl,Keppens}). The inner part of the parallel jet, if composed for instance of
electrons only, creates a magnetic field, which tends to
stabilize the collimation for its outer part (which is the most energetic
part in the precedent example $\rho _{1}=0.3466M$). The strength of the
magnetic field depends on the relative part of the charged particles in the
jet. The radial structure of the jet, in our model, supported by direct
observations \cite{Giroletti}, would require a stability study. However, a
first rough evaluation shows that the ratio $\sqrt{\delta _{1}}/E_{C}\simeq
\sqrt{\delta _{1}}/E$ remains always $\ll 1$ inside the UR jet, which
let us think that any Kelvin-Helmholtz instability is negligible, in
accordance with the results of extensive studies in 2D \cite{Perucho} or in 3D \cite{Perucho1}.

Recent articles discussed the possibility to generate high energy particles by collisions near a BH \cite{Grib} and evoked the interest to consider the Penrose process \cite{Banados}. Our approach can be seen as a contribution to this debate.

\end{document}